\documentclass[a4paper,fleqn,usenatbib]{mnras}
\usepackage{mathptmx}

\usepackage[T1]{fontenc}
\usepackage{ae,aecompl}

\usepackage[latin1]{inputenc}
\usepackage{multirow}
\usepackage{graphicx} 
\usepackage{multicol}
\usepackage{float}
\usepackage{footnote}
\usepackage{amssymb} 	
\usepackage[figuresright]{rotating}
\usepackage{amsmath} 	
\usepackage{url}
\usepackage{rotating}
\usepackage{caption}

\usepackage{subcaption} 
\usepackage[export]{adjustbox}

\title[The mini-halo in the Perseus Cluster at 230-470 MHz]{Deep 230-470 MHz VLA Observations of the Mini-Halo in the Perseus Cluster}
\author[M. Gendron-Marsolais et al.]{M. Gendron-Marsolais$^{1}$\thanks{E-mail: marie-lou@astro.umontreal.ca}, J. Hlavacek-Larrondo$^{1}$, R. J. van Weeren$^{2}$, T. Clarke$^{3}$, 
\newauthor A. C. Fabian$^{4}$, H. T. Intema$^{5}$, G. B. Taylor$^{6}$, K. M. Blundell$^{7}$ and  J. S. Sanders$^{8}$\\
$^{1}$D\'epartement de Physique, Universit\'e de Montr\'eal, Montr\'eal, QC H3C 3J7, Canada\\
$^{2}$Harvard-Smithsonian Center for Astrophysics, 60 Garden Street, Cambridge, MA 02138, USA\\
$^{3}$Naval Research Laboratory, Code 7213, 4555 Overlook Ave. SW, Washington, DC 20375, USA\\
$^{4}$Institute of Astronomy, University of Cambridge, Madingley Road, Cambridge CB3 0HA\\
$^{5}$Leiden Observatory, Leiden University, Niels Bohrweg 2, NL-2333CA, Leiden, The Netherlands\\
$^{6}$Department of Physics and Astronomy, University of New Mexico, Albuquerque, NM 87131, USA\\
$^{7}$University of Oxford, Astrophysics, Keble Road, Oxford OX1 3RH, UK\\
$^{8}$Max-Planck-Institut f\"ur extraterrestrische Physik, 85748 Garching, Germany \\}

\date{Accepted 2017 April 27. Received 2017 April 27; in original form 2017 January 12}

\pubyear{2017}

\begin{document}
\label{firstpage}
\pagerange{\pageref{firstpage}--\pageref{lastpage}}
\maketitle


\begin{abstract}
We present a low-frequency view of the Perseus cluster with new observations from the Karl G. Jansky Very Large Array (JVLA) at 230-470 MHz. The data reveal a multitude of new structures associated with the mini-halo. The mini-halo seems to be influenced both by the AGN activity as well as by the sloshing motion of the cool core cluster's gas. In addition, it has a filamentary structure similar to that seen in radio relics found in merging clusters. We present a detailed description of the data reduction and imaging process of the dataset. The depth and resolution of the observations allow us to conduct for the first time a detailed comparison of the mini-halo structure with the X-ray structure as seen in the Chandra X-ray images. The resulting image shows very clearly that the mini-halo emission is mostly contained behind the western cold front, similar to that predicted by simulations of gas sloshing in galaxy clusters, but fainter emission is also seen beyond, as if particles are leaking out. However, due to the proximity of the Perseus cluster, as well as the quality of the data at low radio frequencies and at X-ray wavelengths, we also find evidence of fine structure. This structure includes several radial radio filaments extending in different directions, a concave radio structure associated with the southern X-ray bay and sharp radio edges that correlate with X-ray edges. Mini-halos are therefore not simply diffuse, uniform radio sources, but rather have a rich variety of complex structures. These results illustrate the high-quality images that can be obtained with the new JVLA at low radio-frequencies, as well as the necessity to obtain deeper, higher-fidelity radio images of mini-halos in clusters to further understand their origin.
\end{abstract}

\begin{keywords}
Galaxies: clusters: individual: Perseus cluster - galaxies: jets - radio continuum: galaxies - X-rays: galaxies: clusters - cooling flows
\end{keywords}

\section{Introduction}

Radio mode feedback in clusters of galaxies is the process by which the energy released by the central active galactic nuclei (AGN) is injected into the intracluster medium (ICM) through turbulence, shocks or sound waves, compensating its radiative losses (e.g. \citealt{birzan_systematic_2004,dunn_investigating_2006,rafferty_feedback-regulated_2006}). The energy source of this mechanism consists of relativistic jets powered by accretion onto the supermassive black hole (SMBH) that inflate bubbles, displacing the ICM and creating regions of depleted X-ray emission. These bubbles, filled with relativistic plasma (i.e., radio lobes), are often discernible at $\sim \text{GHz}$ frequencies.

\begin{figure*}
\centering
\includegraphics[scale=0.50]{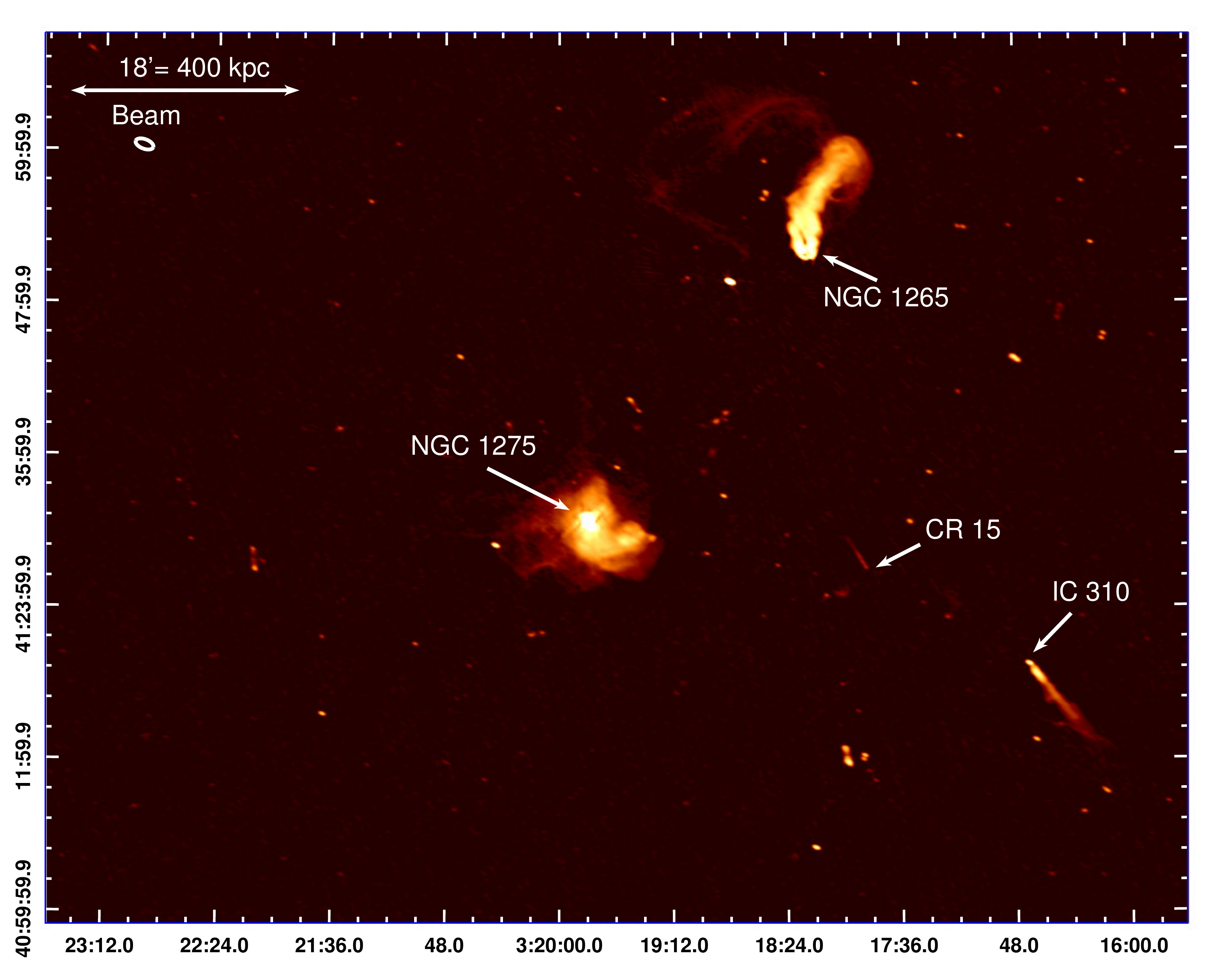}
\caption{The central $2\,^{\circ} \times 1.5\,^{\circ}$ of the total field of view of the JVLA 230-470 MHz radio map obtained in B-configuration. NGC 1275 is the bright source in the middle of the image. Tow wide-angle tail radio galaxies, NGC 1265 (NNW of NGC 1275) and CR 15 (between NGC 1275 and IC 310) as well as IC 310 (WSW of NGC 1275) are clearly visible. The resulting image has a rms noise of 0.35 mJy/beam, a beam size of $22.1 \arcsec \times 11.3 \arcsec$ and a peak of 10.63 Jy/beam.}
\label{fig:image_perseus_JVLA} 
\end{figure*}

The Perseus cluster, the brightest cluster in the X-ray sky, was one of the first examples in which such radio mode feedback was observed. However, in addition to finding two inner cavities filled with radio emission at $5 < r <20$ kpc from the AGN \citep{bohringer_rosat_1993}, another pair of outer cavities was identified further out at $25 < r < 45$ kpc and devoid of high-frequency radio emission \citep{branduardi-raymont_soft_1981,fabian_distribution_1981,churazov_asymmetric_2000}.
These were interpreted as ``ghost" cavities, inflated in the past by the jets of the central SMBH and now rising buoyantly as a gas bubble does in a liquid. As the population of particles filling cavities loses its energy through synchrotron emission, the radio lobes become less clear at higher radio frequencies, providing an explanation of the lack of high-frequency radio emission in the older ghost cavities  (e.g. \citealt{blundell_3c_2002,fabian_properties_2002}).

 \cite{heinz_x-ray_1998} first estimated the input power of the jets originating from the AGN in the central dominant galaxy NGC 1275, known as the Brightest Cluster Galaxy (BCG, \citealt{forman_observations_1972}), to be of the order of $10^{45} \text{ erg s}^{-1}$, a value comparable to the radiative losses. This AGN is thought to be powered by a $ \left( 8^{+7}_{-2} \right)  \times 10^8 \text{M}_{\odot}$ SMBH \citep{scharwachter_kinematics_2013}. 
The advent of the \textit{Chandra} X-ray observatory and its high spatial resolution enabled, in addition, the detection of quasi-spherical ripples interpreted as sound waves in the Perseus cluster \citep{fabian_deep_2003}, 
as well as shocks around cavities \citep{fabian_very_2006},
a semicircular cold front,
two new elliptical cavities interpreted as potential ghost bubbles,
two large regions of weak X-ray luminosity (the northern trough and the southern bay)
and a loop-like structure above a long $\text{H} \alpha$ filament \citep{fabian_wide_2011}.

The Perseus cluster also harbours a mini-halo \citep{soboleva_3c84_1983,pedlar_radio_1990,burns_where_1992,sijbring_radio_1993}, a faint diffuse source of radio emission detected so far in about thirty cool core clusters (\citealt{giacintucci_occurrence_2017}, see \citealt{feretti_clusters_2012} for a review). The X-rays and mini-halo structure of Perseus have been compared in \cite{fabian_wide_2011}. This emission differs from that filling the X-ray cavities, being $\gtrsim 3$ times radio fainter and having a steeper spectral index ($\alpha < -1$ for $S(\nu) \propto \nu^{\alpha}$, where $S$ is the flux density and $\nu$ is the frequency, \citealt{giacintucci_new_2014}). Since the radiative timescale of the electrons is much shorter than the time required for them to reach the extent of the mini-halo, the origin of mini-halos remains unclear. 
Two possible mechanisms have been proposed in the literature to explain the mini-halo emission: it might originate from the reacceleration of pre-existing electrons by turbulence \citep{gitti_modeling_2002,gitti_particle_2004} or from the generation of new particles from inelastic collisions between relativistic cosmic-ray protons and thermal protons (e.g. \citealt{pfrommer_constraining_2004}).
Simulations seem to suggest that turbulence created by sloshing motions of the cold gas in the core region is sufficient to re-accelerate electrons (e.g. \citealt{zuhone_turbulence_2013}). A key prediction of these simulations is that mini-halos should be bounded by cold fronts. 
Observationally, the Hitomi Soft X-ray Spectrometer showed that the line-of-sight velocity dispersions are on the order of $164 \pm 10$ km/s in the 30-60 kpc region around the nucleus of the Perseus cluster. This is sufficient to sustain the synchrotron emission of relativistic electrons population \citep{hitomi_collaboration_quiescent_2016} and sets a limit on the maximum energy density in turbulent motions available \citep{fabian_sound_2017}. The efficiency of the acceleration by turbulence depends however on the assumptions on the spectrum of turbulent motions and on the ICM microphysics \citep{brunetti_challenge_2016}.





In this article, we present new, deep  Karl G. Jansky Very Large Array (JVLA) observations of the Perseus cluster in the P-band (230-470 MHz). 
The resolution and sensitivity of these data provide a detailed and extended view of the mini-halo structure, on which we will focus our analysis. 
The recent update of the facilities with the EVLA project offer new abilities to study this structure.
In Section \ref{JVLA Observations and data reduction}, we present the observations and the data reduction of the JVLA dataset. The results are then presented in Section \ref{Results}. Section \ref{Discussion} discusses the different structures found in the radio observations, comparing Perseus to other clusters and to simulations. Results are summarized in Section \ref{Conclusion}.

We assume a redshift of $z = 0.0183$ for NGC 1275 corresponding to a luminosity distance of 78.4 Mpc, assuming $H_{0} = 69.6 \text{ km s}^{-1} \text{Mpc}^{-1}$, $\Omega_{\rm M} = 0.286$ and $\Omega_{\rm vac} = 0.714$. This corresponds to an angular scale of 22.5 kpc arcmin$^{-1}$.

\section{JVLA Observations and data reduction}\label{JVLA Observations and data reduction}

\subsection{JVLA observations}

We obtained a total of 13 hours in the P-band (230-470 MHz) of the Karl G. Jansky Very Large Array (project $13B-026$): 5 h in the A configuration (2014 May 16); 5 h in the B-configuration (2013 November 24) and 3 h in the D configuration (2014 July 6), which have a synthesized beamwidth of 5.6, 18.5 and 200 arscec respectively at this frequency. The JVLA is fitted with new broadband low frequency receivers. The P-band bandwidth has been widened from 300-340 MHz to 230-470 MHz, increasing significantly the sensitivity of the telescope. 
This article focuses only on the B-configuration data as its resolution probes the faint extended emission of the mini-halo structure in the Perseus cluster. 
Although the D configuration observations would normally probe the total extent of the mini-halo, these data alone are not good enough to produce a reliable map of the extended emission due to the confusion limit associated with the poor resolution of the D configuration. In addition, the D-array observations cannot be easily combined with the B-array observations at sufficiently high-dynamic range since there is little overlap in uv-coverage.
The analysis of the A configuration observations, focussing on the AGN jets, will be presented in Gendron-Marsolais et al. 2017 (in prep.).

\begin{figure}
\includegraphics[width=\columnwidth]{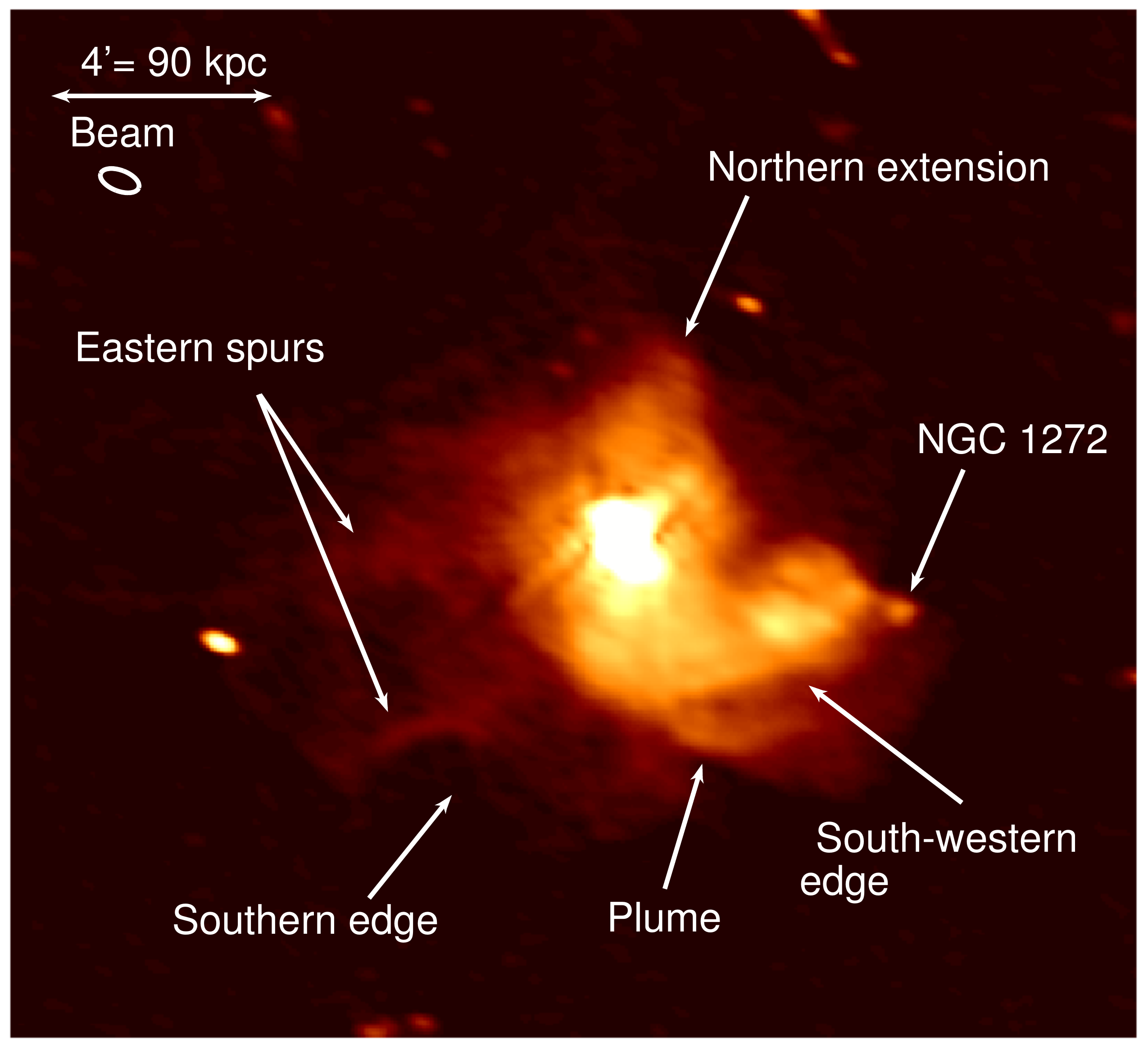}
\caption{A zoom on the emission surrounding NGC 1275 from the 270-430 MHz radio map seen in Fig.~ \ref{fig:image_perseus_JVLA}. The main structures of the mini-halo are identified: the northern extension, the two eastern spurs, the concave edge to the south, the south-western edge and a plume of emission to the south-south-west. The small knob at the end of the western tail is the galaxy NGC 1272.}
\label{fig:image_perseus_JVLA_zoom}
\end{figure}

The JVLA P-band (230-470 MHz) B-configuration data were taken with 27 operational antennas. During the observation period, the operator log reported low band receiver problems with some antennas (9, 11 and 25), which were removed from the dataset at the beginning of the data reduction process. The dataset consists of a total of 58 scans, consisting of about 10 min on 3C48 (for flux, phase and bandpass calibration), another 10 min on 3C147 in the middle of the observation period and a last 10 min on 3C147 at the end (both for phase and bandpass calibration). The rest of the observations consist of scans of 5-10 min each on NGC 1275. The P-band receiver has 16 spectral windows, each comprising of 128 channels with a width of 125 kHz. 

The data reduction was performed with CASA (Common Astronomy Software Applications, version 4.6). A pipeline was specifically developed to account for the strong presence of radio frequency interference (RFI) at low-frequencies and the extremely bright central AGN in Perseus outshining faint structures. The steps of the data reduction process are detailed in the following paragraphs.


The data reduction was performed separately on each spectral window. The data column of the original dataset was therefore split at the beginning into 16 different measurement sets. Early in the use of the new P-band system the polarizations were incorrectly labelled circular instead of linear. The task \textsc{fixlowband} was first applied to correct this issue. Initial flagging was conducted on the two calibrators as well as on the target to remove the most apparent RFI using the mode \textsc{tfcrop} with the task \textsc{flagdata} and the flagger framework \textsc{AOFlagger} \citep{offringa_morphological_2012}.
When those steps of initial flagging had removed most of the RFI, calibration was conducted. Both 3C48 and 3C147 were used as bandpass calibrators using the task \textsc{concat} which concatenates the visibility datasets. By using two bandpass calibrators, we can increase the signal-to-noise ratio on the bandpass solutions.
We inspected visually each calibration table produced with \textsc{plotcal}, identifying and removing outliers. Once each spectral window dataset was calibrated and cleaned of RFI, they were recombined using \textsc{concat}.

The imaging process was performed with a self-calibration method (amplitude and phase), consisting of producing first an image with the task \textsc{clean} and second to derive gain corrections for amplitudes and phases with \textsc{gaincal}, applying these corrections with \textsc{applycal}, producing a new corrected data column. The self-calibration therefore used the target data instead of a calibration source to refine the calibration in an iterative process, producing incremental gain corrections. Once again the tables produced were examined and showed smooth solutions. This procedure was applied three times. Lastly, bandpass and blcal calibrations were conducted to produce the final image. 

Parameters of the clean task had to be carefully adjusted due to the complexity of the structures of Perseus and its high dynamic range (we reached a dynamic range of $30 000$, with an rms of 0.35 mJy/beam and a peak at 10.63 Jy/beam). 
In order to produce a continuum image, a map of the sky-brightness distribution integrated over the frequency range, the frequency response of the interferometer as well as the spectral structure of the radio emission has to be taken into account, specifically with the new generation of broad-band receivers of the JVLA \citep{rao_venkata_parameterized_2010}. Therefore, a multi-scale and multi-frequency synthesis-imaging algorithm (MS-MFS, \citealt{rau_multi-scale_2011}) has been used, choosing the mode \textsc{mfs} in the task \textsc{clean}. To model the frequency dependence of the sky emission, we set the number of Taylor's coefficients to 2 to take into account the complexity of Perseus.
The sky curvature across the wide field of view is corrected by the W-projection algorithm by choosing the \textsc{widefield} grid mode \citep{cornwell_noncoplanar_2008}. The number of w-planes used was set to 480.
We tested the Brigg's robustness parameter \textsc{robust} of the weighting, testing the values $-2$ (uniform), 0 (default) and $+2$ (natural). The default value, 0, gave the best image.
The size of the image ($6144 \text{pixels} \times 6144 \text{pixels} \sim 5.12\,^{\circ} \times 5.12\,^{\circ}$, while the full width at half power of the field of view in the middle of P band is around $2.4\,^{\circ}$) was chosen to be big enough to include all bright sources surrounding NGC 1275.
Considering the synthesized beamwidth of the B-configuration at 230-470 MHz ($18.5 \arcsec$) the chosen cell size was $3 \arcsec \times 3 \arcsec$.
A multi-scale cleaning algorithm \citep{cornwell_multiscale_2008} was used in order to take account of the different scales of the structures we were imaging: the point source ($0 \arcsec$), the inner cavities ($15 \arcsec = 5 \text{ pix}$), the mini-halo ($30 \arcsec = 10 \text{ pix}$), the ghost cavities ($60 \arcsec = 20 \text{ pix}$) and NGC 1265 ($150 \arcsec = 50 \text{ pix}$).
Most of the \textsc{clean} tasks were run in interactive mode in order to control the number of iterations as well as to build a cleaning mask interactively. 
We also used the tool \textsc{PyBDSM} (Python Blob Detection and Source Measurement software, \citealt{mohan_pybdsm:_2015}) to build the first version of the mask.
The number of iterations for each \textsc{clean} task during the self-calibration process was about 100 000.

In addition to the automatic flagging, manual flagging was done before the self-calibration on the target, examining the amplitude vs. the UV-wave with \textsc{plotms} allowing to identify bad channels, baselines or time ranges. 
To identify sources of artifacts during the imaging process, we use the parameter \textsc{selectdata} of the task \textsc{clean}, imaging all data except one antenna and iterating over antennas. This method was applied similarly for scans and spectral windows. It allows the identification of one bad antenna (antenna 23). 
Inspection with \textsc{plotms} of the corrected real vs. imaginary portions of the visibilities indicated no more bad data or outliers.
The final image resulting from this data reduction process is presented in Figs.~ \ref{fig:image_perseus_JVLA} and \ref{fig:image_perseus_JVLA_zoom}.

\subsection{X-ray observations}

We use the final composite fractional residual image from \cite{fabian_wide_2011} to compare our radio data to the X-ray emission. 
This image is the result of a total of 1.4 Ms \textit{Chandra} observations, 900 ks of ACIS-S data \citep{fabian_chandra_2000,fabian_deep_2003,fabian_very_2006} of the central $180 \times 180 \text{kpc}$ combined with 500 ks of ACIS-I large-scale observations. 
The data have been cleaned from flares, reprocessed, reprojected to match coordinates, merged together and exposure map corrected (see details in \citealt{fabian_very_2006,fabian_wide_2011}).
This image was then adaptively smoothed with a top-hat kernel with the bin size chosen to contain 225 counts. Ellipses were fitted in this smoothed image to logarithmic equally-spaced levels of surface brightness. A model was constructed by interpolating between the elliptical contours (in log surface brightness level). The final image, shown in Fig.~ \ref{fig:image_perseus_xray_analysis}- left, is the fractional difference between the adaptively smoothed image and this model.
Furthermore, to examine the position and shape of the western cold front found in Perseus \citep{fabian_wide_2011} we use a Gaussian gradient magnitude (GGM) filtered image (see Fig.~ \ref{fig:image_perseus_xray_analysis}- middle), highlighting the edges in the merged X-ray image \citep{sanders_detecting_2016}. 
Finally, we use the temperature map from \cite{fabian_wide_2011} generated with the \textit{Chandra} observations binned into regions of 22 500 counts each (see Fig.~ \ref{fig:image_perseus_xray_analysis}- right). The errors associated with the temperature map values vary from $\sim 1\%$ to $6\%$.

\subsection{Previous radio observations at 74 MHz, 235 MHz and 610 MHz}

We use previous radio observations of the Perseus cluster at 74 MHz, 235 MHz and 610 MHz to complete our study of the mini-halo. 
We use archival VLA data at 74 MHz (A configuration, see Fig.~ \ref{fig:contours} top-left, \citealt{blundell_3c_2002}).
We also present for the first time new Giant Metrewave Radio Telescope (GMRT) observations of the Perseus cluster at 235 MHz (PI: Blundell, see Fig.~ \ref{fig:contours} top-right). A total of 10 hours of observations were obtained. The GMRT data were reduced with the standard astronomical image processing system (aips, version 31DEC11, Greisen 2003), following the normal procedure (RFI removal, calibration and imaging).
Several rounds of flagging, including manual flagging were applied, resulting in more than 50 per cent of the data being flagged. The data were also calibrated in phase and amplitude, and 3C 286 and 3C 48 were used as flux calibrators. However, due to the dynamic range limits of the GMRT, we were only able to reach a noise level of 10 mJy/beam (beam size of $13\arcsec \times 13 \arcsec$), which is significantly higher than the JVLA noise level. We therefore do not discuss these data in detail.
We also use Westerbork Synthesis Radio Telescope (WSRT) 610 MHz contours from \cite{sijbring_radio_1993} (see Fig.~ \ref{fig:contours} bottom-right).

\subsection{Optical observations}

\begin{figure*}
    \centering
    \begin{subfigure}[b]{0.45\textwidth}
    \includegraphics[width=\textwidth]{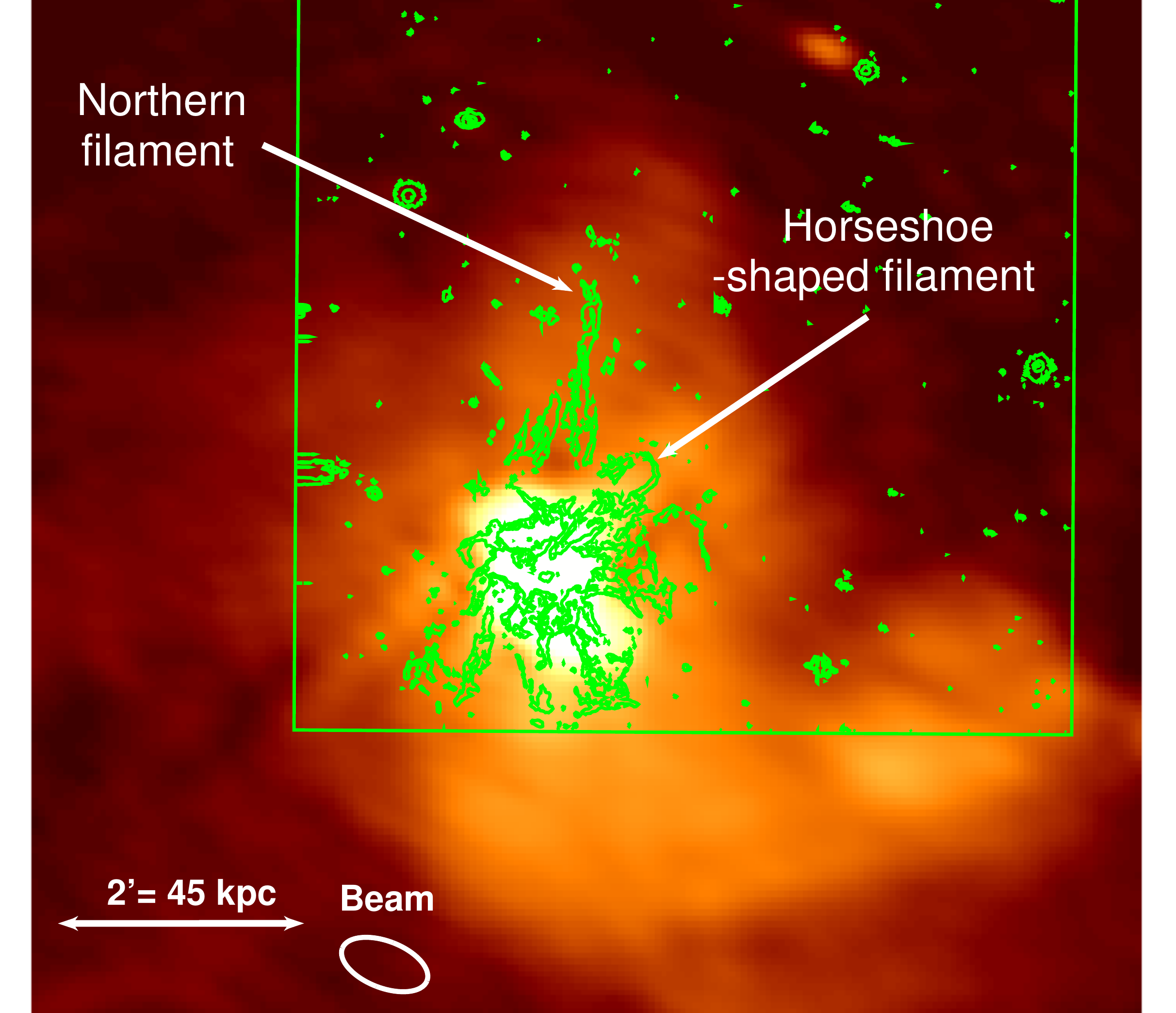}
    \end{subfigure}
    \begin{subfigure}[b]{0.45\textwidth}
    \includegraphics[width=\textwidth]{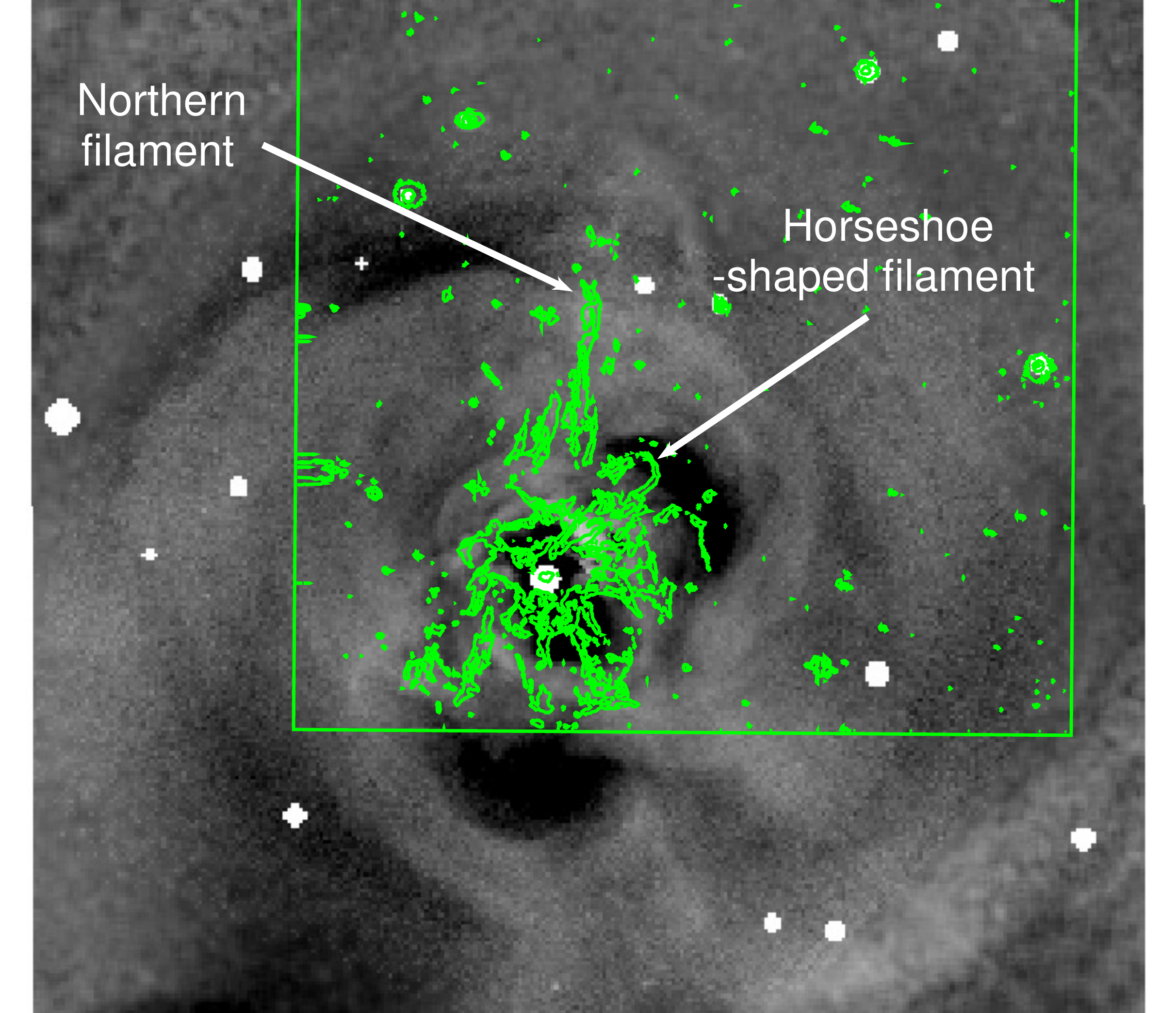}
    \end{subfigure}
\caption{Left - JVLA 270-430 MHz B-configuration image with $\text{H} \alpha$ contours (in green) of the continuum-subtracted $\text{H} \alpha$ map from \protect\cite{conselice_nature_2001}. The green region indicates the edge of the $\text{H} \alpha$ image. The horseshoe-shaped filament and the northern filament are identified. Right - The fractional residual X-ray image with the same $\text{H} \alpha$ contours.}
\label{fig:image_perseus_Halpha}
\end{figure*}

The BCG NGC 1275 is surrounded by a giant filamentary $\text{H} \alpha$ nebula extending over 100 kpc, made of two components: a high-velocity system (8200 km s$^{-1}$), corresponding to a disrupted foreground galaxy, and a low-velocity system (5265 km s$^{-1}$). In order to make a detailed comparison of this rich nebula with the radio and X-ray emission morphology, we use the continuum-subtracted $\text{H} \alpha$ map from \cite{conselice_nature_2001} produced with the Wisconsin-Indiana-Yale-NOAO (WIYN) telescope observations (see Fig.~ \ref{fig:image_perseus_Halpha}).

\begin{figure*}
\centering
    \begin{subfigure}[c]{0.33\textwidth}
    \vspace{-0pt}
    \includegraphics[width=\textwidth]{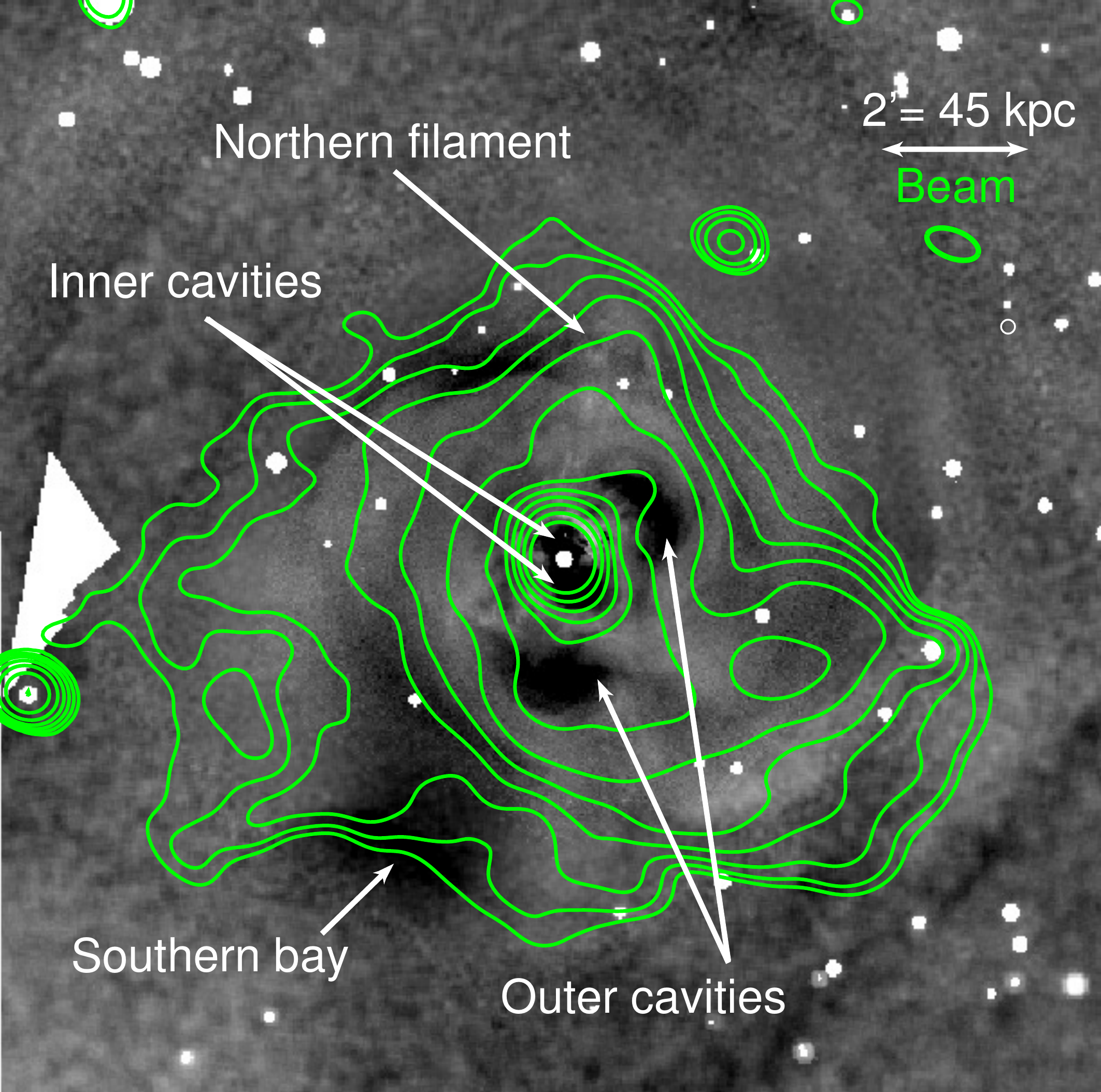}
    \vspace{8pt}
    \end{subfigure}
    \begin{subfigure}[c]{0.33\textwidth}
    \vspace{-0pt}
    \includegraphics[width=\textwidth]{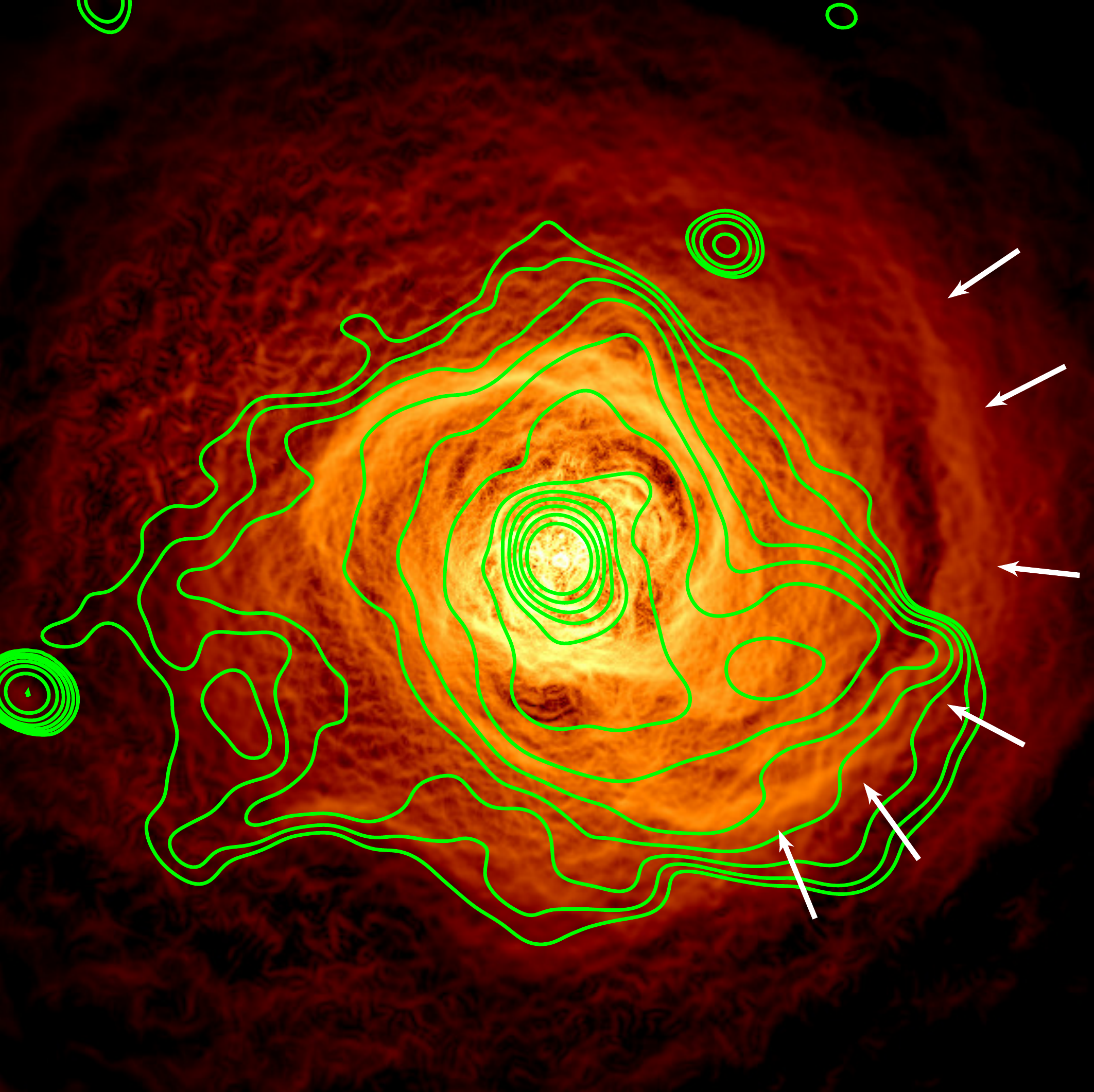}
    \vspace{8pt}
    \end{subfigure}
    \begin{subfigure}[c]{0.33\textwidth}
    \vspace{-0pt}
    \includegraphics[width=\textwidth]{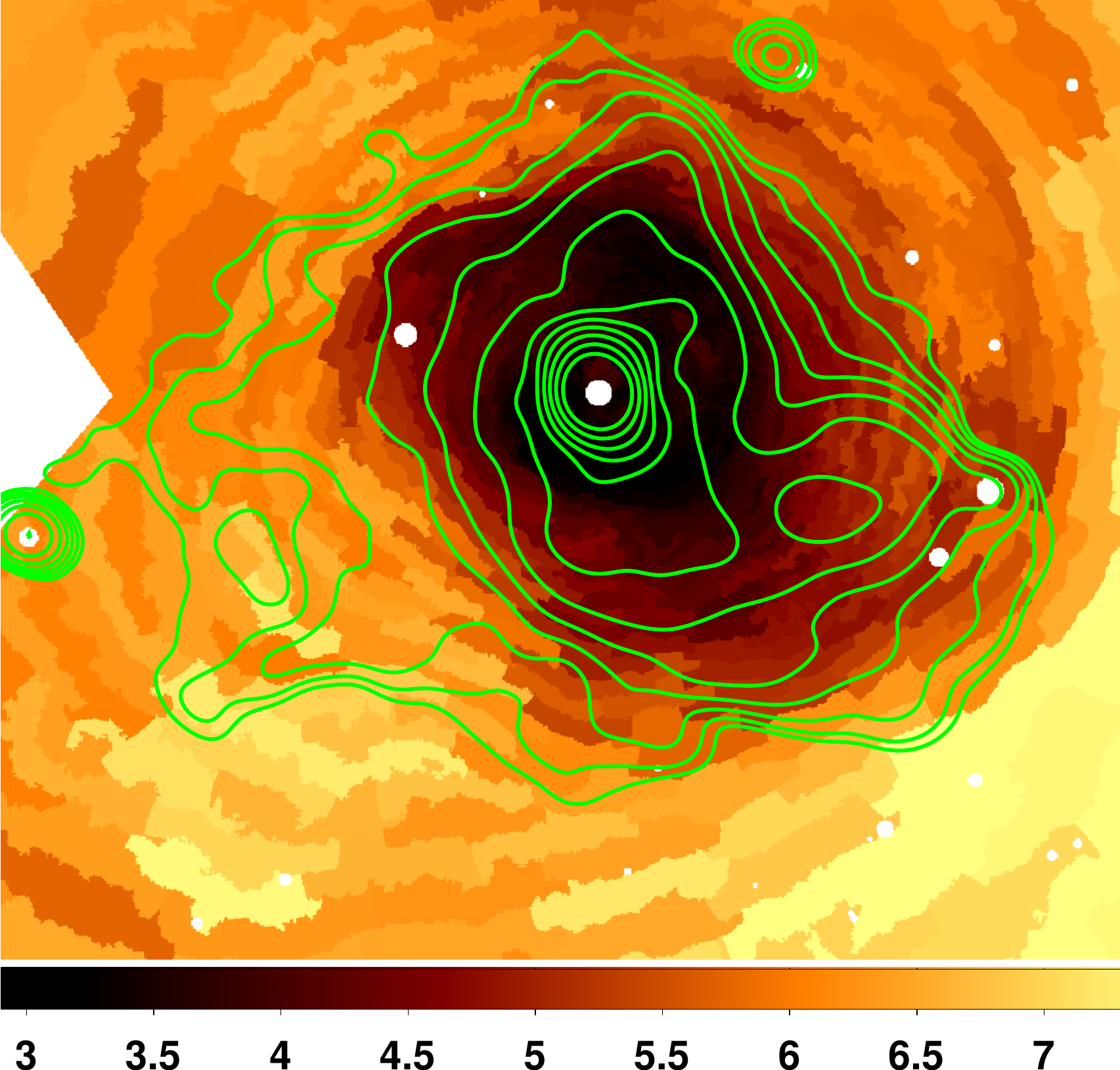}
    \vspace{8pt}
    \end{subfigure}
\caption{Left - \textit{Chandra} final composite fractional residual image from \protect\cite{fabian_wide_2011} in the 0.5-7 keV band (total of 1.4 Ms exposure) with 270-430 MHz contours from $5\sigma = 1.75$ mJy/beam to 1 Jy overlaid from JVLA B-configuration. Middle - GGM filtered image of the merged X-ray observations with Gaussian width $\sigma=4$ pixels \citep{sanders_detecting_2016} with the same 270-430 MHz JVLA contours. The position of the western cold front is indicated with white arrows. Right - Central part of the temperature map of the Perseus cluster from \protect\cite{fabian_wide_2011} with signal-to-noise ratio of 150 with the same 270-430 MHz JVLA contours. Units are keV.}
\label{fig:image_perseus_xray_analysis}
\end{figure*}

\section{Results}\label{Results}

Fig.~\ref{fig:image_perseus_JVLA} shows the central part of the B-configuration final map obtained from the data reduction and imaging process described in Section \ref{JVLA Observations and data reduction}. 
The resulting image has a rms of 0.35 mJy/beam, peak at 10.63 Jy/beam and a beam size of $22.1'' \times 11.3''$.
Beyond the central emission from NGC 1275, the large field of view includes NGC 1265, a wide-angle tail radio galaxies, and IC 310, an active radio galaxy with a blazar-like behavior and jets observed at a viewing angle of $10\,^{\circ} - 20\,^{\circ}$ \citep{aleksic_rapid_2014,ahnen_first_2017}, both discovered by \cite{ryle_radio_1968}.
The analysis of their complex morphologies will be presented in future work (Gendron-Marsolais et al. 2017 in prep.).
The smaller head-tail source CR 15 is also found between NGC 1275 and IC 310, the tail pointing in a northeast direction \citep{miley_active_1972}.

Fig.~\ref{fig:image_perseus_JVLA_zoom} shows a zoom of the central radio emission surrounding NGC 1275. 
Features running roughly north-west/south-east through the core are the remaining artifacts due to some problematic antennas all located in the same arm of the JVLA. Some of these antennas were removed, but removing all would have caused the beam to be extremely elongated. Since the artifacts are mostly located near the central AGN, we optimized the removal of the antennas such that the artifacts would be minimal while allowing the beam to remain roughly circular.
The diffuse mini-halo structure extends up to $\sim 150$ kpc from the AGN and shows a complex structure. 
The general shape of the mini-halo in the Perseus cluster has an irregular morphology, curving counterclockwise.
It is also elongated in the direction of the radio bubble system.
In addition to this large-scale structure shape, several fine structure details in the emission have been identified: two spurs are seen to the east and southeast of the AGN, an extension to the north, an edge to the south-west, a plume of emission to the south-south-west and a concave edge to the south. 
An analysis of these structures is presented in the following section.

As in \cite{giacintucci_new_2014}, we estimate the average radius of the mini-halo as $R=\sqrt{R_{\rm max}\times R_{\rm min}}$ based on the $3\sigma$ contours. With $R_{max} = 7 \arcmin = 157.5\text{ kpc}$ and $R_{\rm min} = 3.5 \arcmin = 78.75 \text{ kpc}$, this gives 111 kpc, which places Perseus's mini-halo as average in terms of size.
The true size of the mini-halo could, however, be larger due to the fact that larger structures are resolved out in the B-array observations. 
We also measured its total flux density using the CASA task  \textsc{imstat}. Integrating the flux density in an annulus centered on the AGN from $1 \arcmin$ (roughly corresponding to a 0.1 Jy contour) up to $R_{\rm max}$ (excluding the central AGN contribution) gives $13.0 \pm 0.7$ Jy.

\begin{figure*}
 \centering
 \begin{subfigure}[b]{0.42\textwidth}
     \caption*{\large{74 MHz VLA contours}}
     \vspace{-2mm}
     \includegraphics[width=\textwidth]{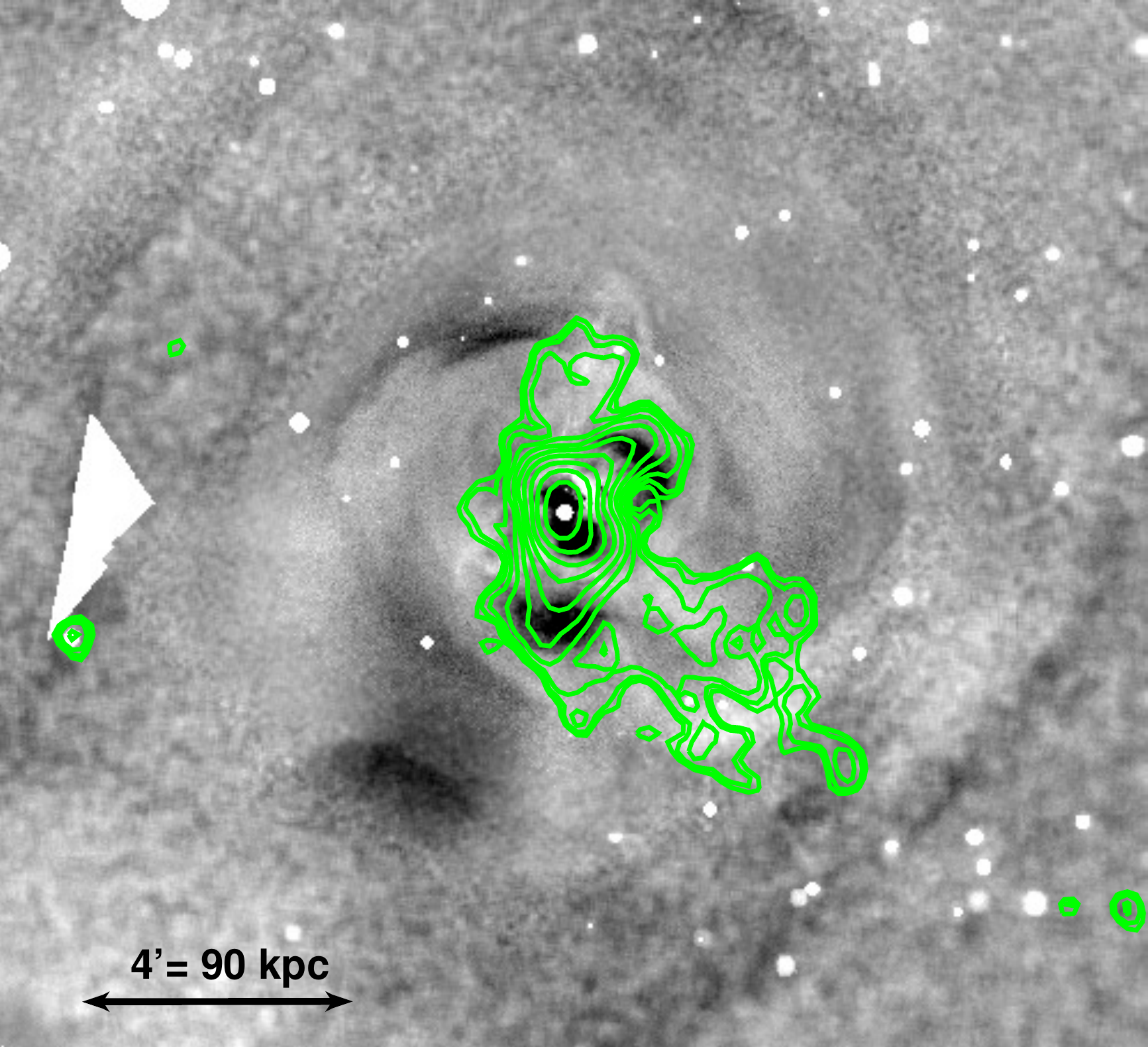}
     \vspace{2pt}
 \end{subfigure}
     \quad
 \begin{subfigure}[b]{0.42\textwidth}
      \caption*{\large{235 MHz GMRT contours}}
      \vspace{-2mm}
      \includegraphics[width=\textwidth]{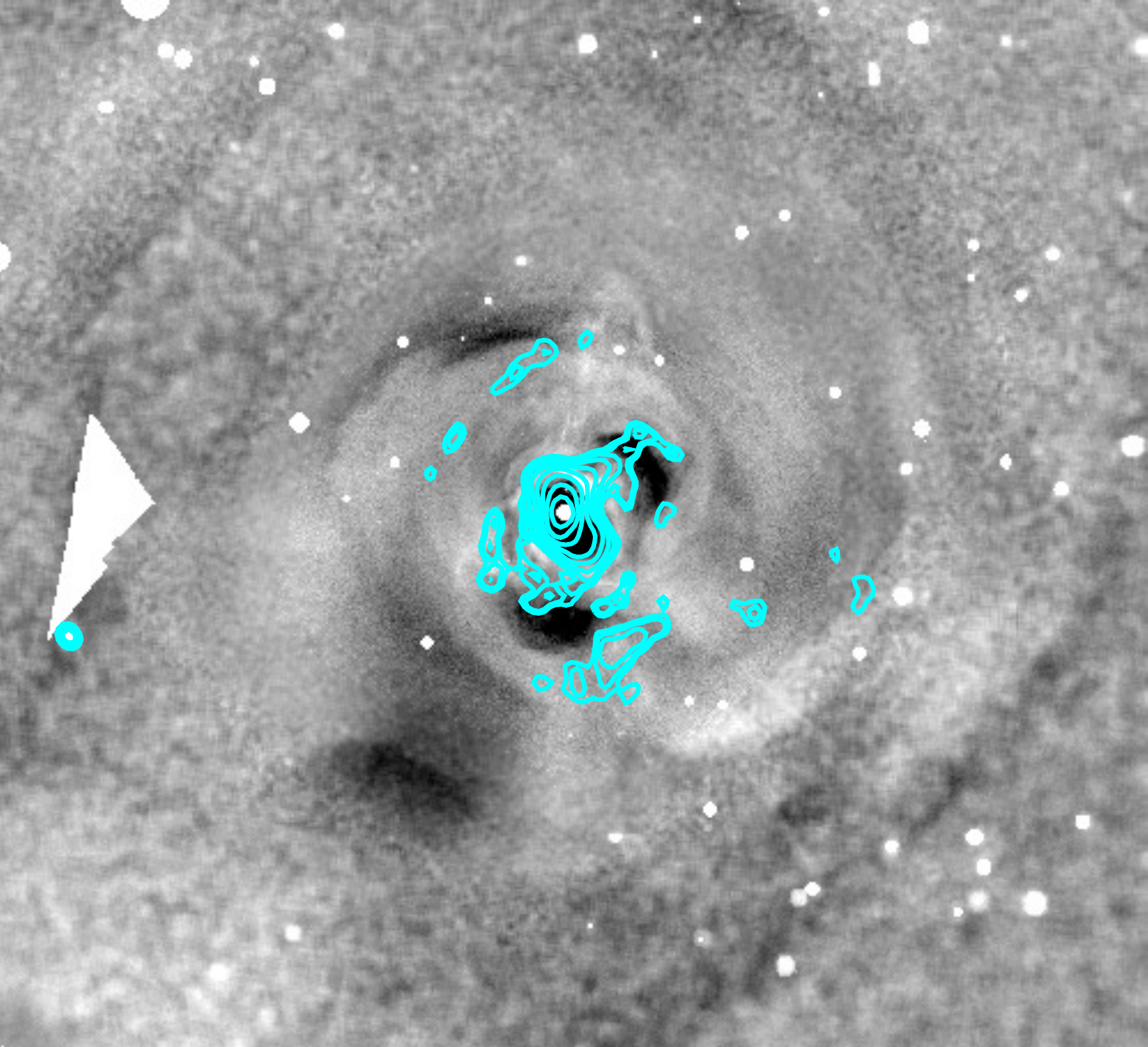}
      \vspace{2pt}
 \end{subfigure}

 \begin{subfigure}[b]{0.42\textwidth}
      \caption*{\large{270-430 MHz JVLA contours}}
      \vspace{-2mm}
      \includegraphics[width=\textwidth]{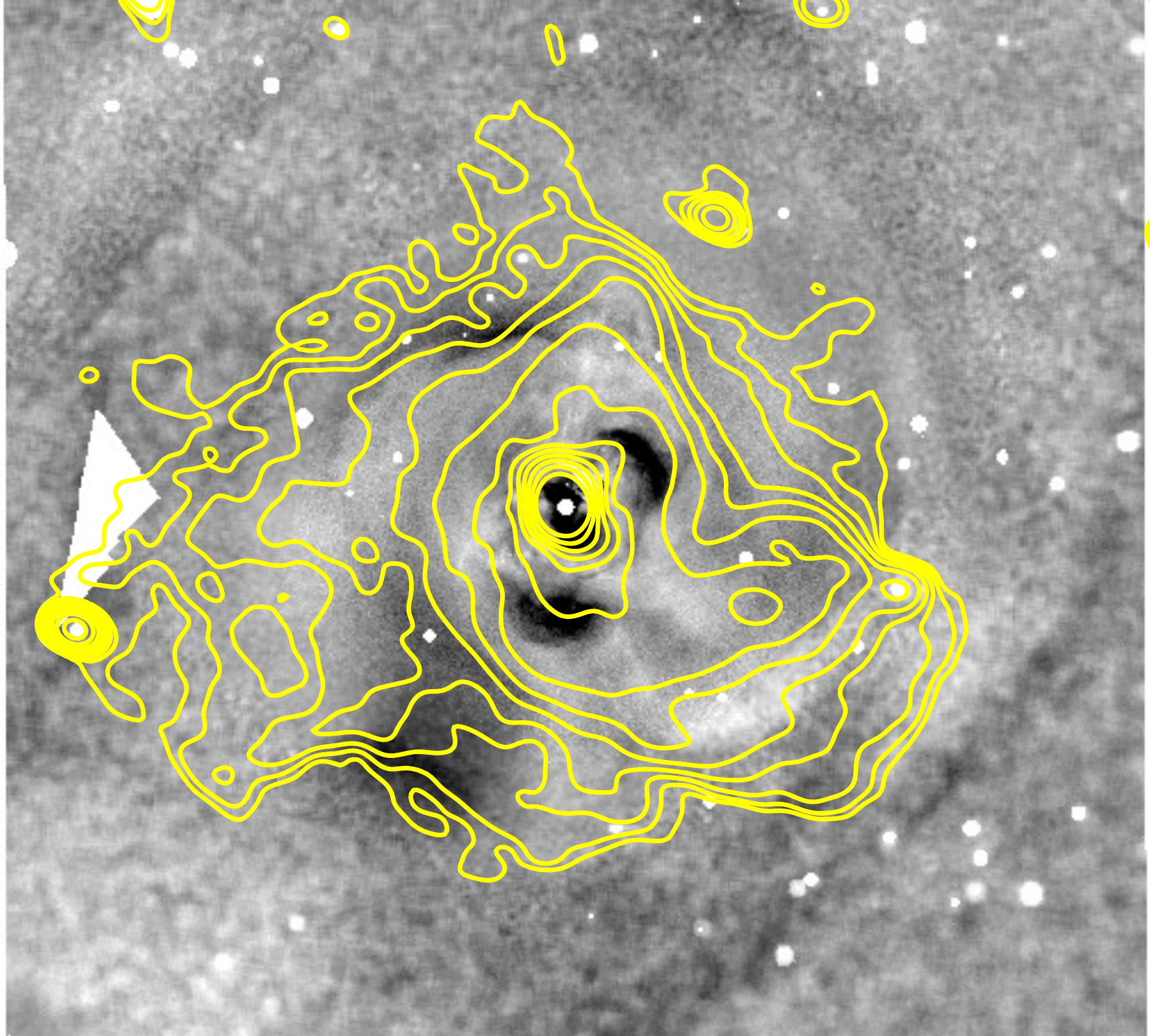}
 \end{subfigure}
 \quad
 \begin{subfigure}[b]{0.42\textwidth}
      \caption*{\large{610 MHz WSRT contours}}
      \vspace{-2mm}
      \includegraphics[width=\textwidth]{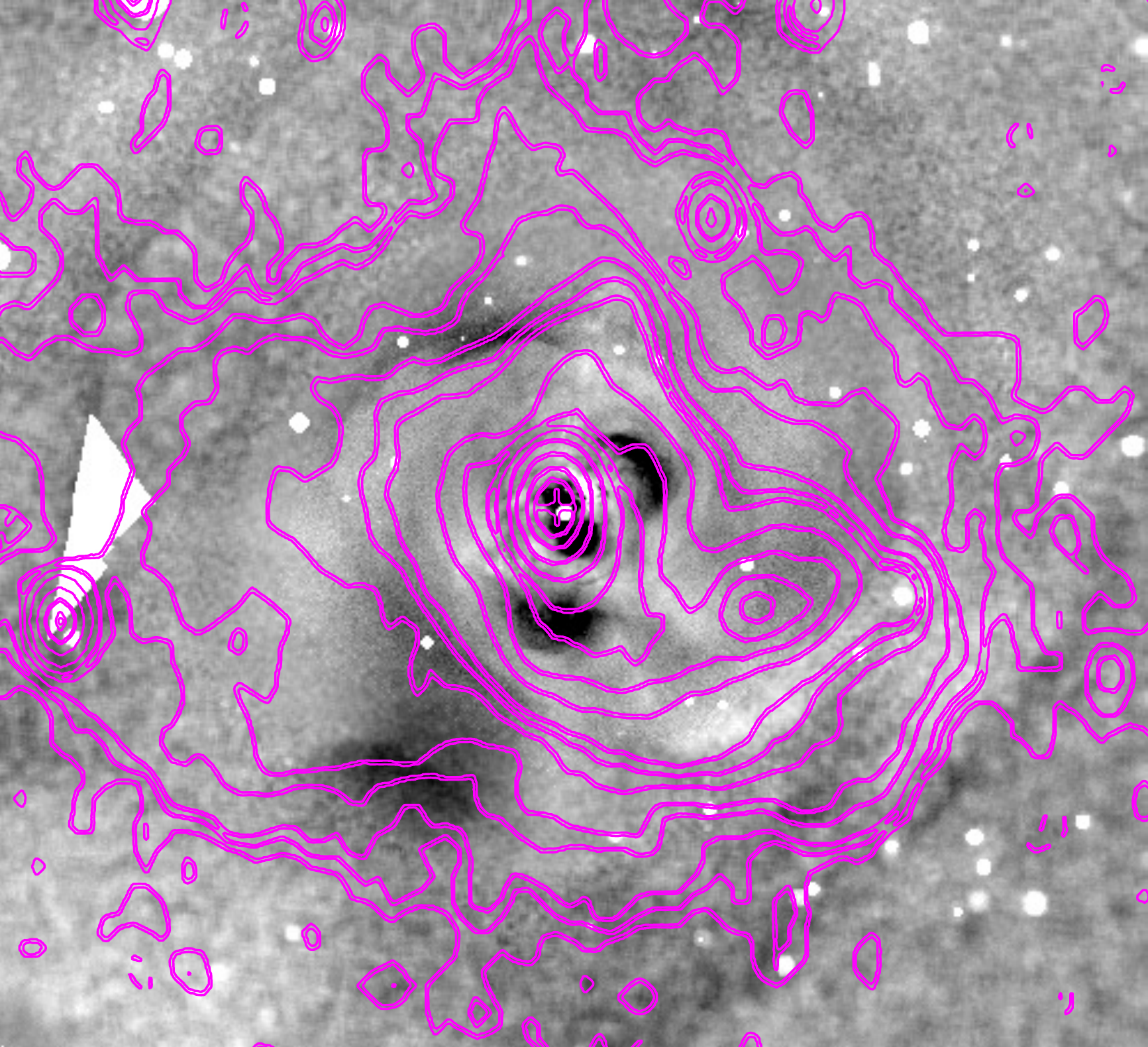}
 \end{subfigure}
 \caption{\textit{Chandra} final composite fractional residual image from \protect\cite{fabian_wide_2011} in the 0.5-7 keV band (total of 1.4 Ms exposure) with radio contours at different frequencies overlaid.
 Top-left: 74 MHz A configuration VLA contours (synthesized beamwidth of $24\arcsec$, $\sigma_{\rm rms} = 80 \text{ mJy/beam}$). A total of 11 contours are drawn, increasing logarithmically from $0.3 \text{ Jy/beam}$ to $36.2 \text{ Jy/beam}$ \protect\citep{blundell_3c_2002}. 
 Top-right: 235 MHz GMRT contours (synthesized beamwidth of $13\arcsec$). 10 contours are drawn, increasing logarithmically from $5\sigma_{\rm rms} = 50 \text{ mJy/beam}$ to $9\text{ Jy/beam}$. 
 Bottom-left: 270-430 MHz contours from the new JVLA B-configuration (beamwidth of $22.1'' \times 11.3''$). A total of 13 contours are drawn, also increasing logarithmically from $3\sigma = 1.05$ mJy/beam to 1 Jy. 
 Bottom-right: 610 MHz WSRT contours from \protect\cite{sijbring_radio_1993}, synthesized beamwidth of $29 \arcsec \times 44 \arcsec$ and $\sigma_{\rm rms} = 0.4 \text{ mJy/beam}$). The contours levels are -0.8 (dashed), 0.8, 1.6, 2.4, 5, 7.5, 15, 22.5, 30, 60, 90, 120, 150, 300, 850, 2500, 5000 and 10000 \text{mJy/beam}.
 }\label{fig:contours}
\end{figure*}

\section{Discussion}\label{Discussion}

\subsection{Comparison with previous radio observations}

Fig.~\ref{fig:contours} shows the 74 MHz, 235 MHz, 270-430 MHz and 610 MHz radio contours overlaid on the fractional residual X-ray image. The mini-halo is seen clearly at 270-430 MHz and 610 MHz, while only spurs of radio emission extend outside the inner and outer X-ray cavities at 74 MHz and almost no emission is seen outside the inner cavities at 235 MHz with the GMRT image at these sensitivity levels. Mini-halos generally have steep spectral indexes but the noise levels of the 74 and 235 MHz images ($80$ and $10 \text{mJy/beam}$, respectively) are too high for the faint mini-halo to be detected. 
The new JVLA facilities have produced an order of magnitude deeper image than
the previous 330 MHz VLA data ($\sigma_{\rm rms} = 7 \text{mJy/beam}$), with high-fidelity, allowing the detection of the mini-halo emission to much larger radii and in much finer detail.
The northern extension of emission as well as hints of the presence of the southern and south-western edges were already visible in 610 MHz WSRT contours from \cite{sijbring_radio_1993}.
The synthesized beam size of these observations ($29\arcsec \times 44 \arcsec$) being about five times larger than the beam size of our 270-430 MHz observations ($22.1\arcsec \times 11.3 \arcsec$), it only probed blurred emission from the eastern spurs, the plume and the southern edge, not fine filaments or sharp edges compared to the new JVLA observations. Our higher resolution image therefore allows us for the first time to study the structure of this mini-halo in detail.

\subsection{Large-scale structure of the mini-halo}

In order to understand the origin of the mini-halo emission, it is useful to compare its morphology with images from other wavelengths. This approach led to the first reports of mini-halo - sloshing cold front correspondences in \cite{mazzotta_radio_2008} for the relaxed galaxy clusters RX J1720.1+2638 and MS 1455.0+2232.
The systematic search for mini-halos in clusters by \cite{giacintucci_new_2014} has shown indications of gas sloshing in the X-ray observations of most of the 21 clusters with mini-halos. The mini-halos were contained inside the sloshing region in many of them.
These correspondences support the reacceleration hypothesis according to which cooled relativistic electrons injected by past AGN activity are reaccelerated by turbulence that may be produced by sloshing motion \citep{gitti_modeling_2002,gitti_particle_2004, mazzotta_radio_2008,zuhone_turbulence_2013}.
The nearby and bright Perseus cluster offers an opportunity to study this correlation as its proximity gives the sensitivity and resolution needed to probe the details of the mini-halo as well as the cold front structures.
The B-configuration radio contours starting at $5\sigma$ are overlaid on \textit{Chandra} X-ray image \citep{fabian_wide_2011} on Fig.~\ref{fig:image_perseus_xray_analysis}- left. The position of the inner and outer cavities are indicated on the figure. One of the most striking features in the deep X-ray images of the Perseus cluster is the spiral pattern of the X-ray emitting gas. A similar trend is also clearly seen in the temperature map (\citealt{fabian_very_2006}, see Fig.~ \ref{fig:image_perseus_xray_analysis}- right) and the western part of it was identified as a cold front \citep{fabian_wide_2011}, a sharp contact discontinuity between gas regions with different temperatures and densities. 
The characteristic spiral pattern of cold fronts are created by the sloshing of gas in a gravitational potential perturbed by a minor merger (see \citealt{markevitch_shocks_2007} for a review). High-resolution simulations of cluster mergers also show that cold fronts are produced by minor mergers and can persist over gigayear time scales \citep{ascasibar_origin_2006}. In the case of the Perseus cluster, a chain of bright galaxies visible to the west of the BCG NGC 1275 have been identified as the possible source of disturbance \citep{churazov_xmm-newton_2003}. Interestingly, the curving shape of the mini-halo seems to be well aligned with the sloshing pattern. It matches both the size and the direction of curvature (counterclockwise). However, the mini-halo is also elongated in the direction of the cavity system.
This spatial correlation is consistent with the scenario that AGN feedback could contribute to the injection of turbulence in the ICM and reaccelerate the relativistic particles responsible for the mini-halo emission \citep{cassano_morphological_2008}.

Fig.~\ref{fig:image_perseus_xray_analysis}- middle shows the X-ray GGM filtered image with the same radio contours. The position of the western cold front is indicated by the arrows. This image shows very clearly how the mini-halo emission is mostly contained behind the cold front: there is a sharp edge in the radio image associated with the mini-halo, but the particles appear as well to leak out as there is an even fainter (2 to 3 times fainter) part of the mini-halo that extends beyond the cold front in the south-western direction. Fig.~\ref{fig:image_perseus_xray_analysis}- right shows the central $13\arcmin$ of the temperature map from \cite{fabian_wide_2011} compared with the mini-halo emission. As in the GGM filtered image, the emission is mostly bounded by the western cold front.

Another interesting large scale structure present in the mini-halo of the Perseus cluster is the southern edge identified in Fig.~\ref{fig:image_perseus_JVLA_zoom}. Compared with the X-ray observations (see Fig.~ \ref{fig:image_perseus_xray_analysis}- left), the radio emission seems to avoid the southern bay, an intriguing feature located about 100 kpc south of the nucleus, first reported in \cite{fabian_very_2006}. 
Recently, \citealt{walker_is_2017} investigates these 'bay' structures, found in three nearby relaxed clusters: Perseus, Centaurus and Abell 1795. Bays behave like cold fronts but have the opposite curvature toward the interior of the cluster. According to simulations of gas sloshing, they might be resulting from Kelvin-Helmholtz instabilities. We refer the reader to \citealt{walker_is_2017} for more details.

\subsection{Filamentary structure}

Very few filamentary structures like the ones present in the P-band JVLA observations of the Perseus cluster have been observed before in mini-halos. 
In Abell 2626, two elongated, $\sim 5\text{ kpc}$ thick, arc-like radio features with longitudinal extensions of $\sim 70$ kpc are detected in its mini-halo \citep{gitti_particle_2004,gitti_puzzling_2013}.
In the case of Perseus, the two eastern spurs identified in Fig.~\ref{fig:image_perseus_JVLA_zoom} are $\geq 10\text{ kpc}$ thick and extend over $\sim 150\text{ kpc}$ in scale. Interestingly, similar filaments are found in radio relics (e.g. the relic in Abell 2256, \citealt{owen_wideband_2014}), large elongated diffuse polarized radio sources located at cluster peripheries. Relics could result from synchrotron emission of electrons reaccelerated by mergers or accretion shocks \citep{ensslin_cluster_1998,brunetti_cosmic_2014,van_weeren_case_2017}.
In Perseus though, no shocks corresponding with the position of the filaments are known.
We can also speculate that these filaments trace regions of enhanced magnetic fields or locally enhanced turbulence. Alternatively, they could reflect the original distribution of fossil plasma, for example from an old AGN outburst (up to a Gyr ago) that are re-accelerated by turbulence or weak shocks.

As shown in Fig.~\ref{fig:image_perseus_Halpha}, the northern extension of the mini-halo also matches the position of the northern filament seen in the $\text{H} \alpha$ map from \cite{conselice_nature_2001}, a long ($\sim 45$ kpc) and thin filament part of the large filamentary nebula surrounding NGC 1275. 
The loop-like X-ray structure extending at the end of the northern filament has been interpreted as fallback gas dragged out to the north by previously formed bubbles \citep{fabian_wide_2011}. As for the elongated shape of the mini-halo aligned with the cavity system, the northern extension shows another correlation of the mini-halo with the relativistic jets. Therefore, the shape of mini-halos seems to originate both from sloshing and from past AGN activity.

\subsection{Qualitative comparison with simulations}

The quality of both the radio and X-ray data allows the comparison of our observations with high-resolution magnetohydrodynamic simulations of gas sloshing in galaxy clusters for example as in \cite{zuhone_turbulence_2013}.
Additionally, the clear spiral pattern seen in X-ray observations and temperature map suggests that the plane of the sloshing pattern is perpendicular to our line of sight. This inclination allows us to directly compare the observations the Perseus cluster with the $z$-projections in \cite{zuhone_turbulence_2013}. The authors show projected gas temperature maps at several epochs with 153, 327, and 1420 MHz radio contours overlaid. The general shape and extent of Perseus's mini-halo most closely resembles the central part of the simulated radio contours .
However, the simulated radio observations also show a patchy tail of emission, absent from the P-band JVLA observations of the mini-halo.

\subsection{Implications for our understanding of mini-halos}

The deep JVLA observations of the Perseus cluster combined with the cluster's properties (proximity, brightness and sloshing plane inclination) offers a unique opportunity to study mini-halo structures. These low-frequency observations have revealed lots of structures, unlike the present observations of most mini-halos which appear to be of fuzzy and uniform emission.
This could be due to the resolution and sensitivity of the radio observations of mini-halos.
Few mini-halos also present structures, e.g. the arc-like radio features in Abell 2626 \citep{gitti_particle_2004,gitti_puzzling_2013} or the spiral-shaped tail of emission in RX J1720.1+2638 with a length of $\sim 230$ kpc \citep{giacintucci_mapping_2014}.
Perseus's mini-halo has similar size but a much higher flux density, 3 to 4 orders of magnitude higher than other mini-halos \citep{giacintucci_new_2014}.
Even with this level of detail, the emission is still mostly constrained behind the sloshing cold front, delimited by a sharp radio edge, providing a qualitative test of the reacceleration hypothesis. However, faint emission is also seen beyond this edge as if particles appear to leak out (see Fig.~\ref{fig:image_perseus_xray_analysis}- middle).
Again, as larger structures are resolved out in the B-array observations,  we could still miss a large-scale  diffuse mini-halo that would extend well beyond the cold front edge indicating that there must be a source of acceleration that goes beyond the cold front.
\section{Conclusion}\label{Conclusion}

The Perseus cluster is a fantastic laboratory to study all processes taking place in a typical cluster as it is internally perturbed by the nuclear outburst of the cluster's brightest galaxy NGC 1275 active galactic nuclei, as well as externally affected by its interaction with its surrounding environment. We present a detailed radio map of the Perseus cluster obtained from 5 h of observations with the JVLA at 230-470 MHz in the B-configuration. 
A CASA pipeline has been specifically developed to reduce this dataset, taking into account the high dynamic range and the multi-scale nature of the Perseus cluster, as well as the strong presence of RFI. 
In summary, we conclude the following.

\begin{enumerate}
\item This work has provided an extended low-frequency view of the mini-halo in the Perseus cluster. Several structures have been identified: the northern extension, two filamentary spurs to the east and a clear edge avoiding the X-ray southern bay. The general shape of the mini-halo is curving counterclockwise and is elongated in the direction of the cavity system. At 230-470 MHz, Perseus's mini-halo extends up to 135 kpc from the nucleus and has a total flux density of 12.64 Jy.
\item The comparison of the 230-470 MHz map with deep \textit{Chandra} observations has shown that the mini-halo is enclosed mostly behind the western sloshing cold front, qualitatively supporting the reacceleration hypothesis. However, fainter emission is also seen beyond, as if particles leaking out.
\item The large-scale and fine structure show a correlation of the mini-halo emission with both the sloshing motion and the relativistic jets of the AGN. 
\item Mysterious filamentary spurs of emission are found to the east, similar to radio relics, but no shocks corresponding with the position of the filaments are known.
\item The shape of the mini-halo resembles the central simulated synchrotron radiation in magnetohydrodynamic simulations of gas sloshing in galaxy clusters for example from \cite{zuhone_turbulence_2013}.
\end{enumerate}

These results demonstrate the sensitivity of the new JVLA, as well as the necessity to obtain deeper, higher-fidelity radio images of mini-halos in clusters to further understand their origin.

\section*{Acknowledgments}
MLGM is supported by NSERC through the NSERC Postgraduate Scholarships-Doctoral Program (PGS D) and Universit\'e de Montr\'eal physics department.
JHL is supported by NSERC through the discovery grant and Canada Research Chair programs, as well as FRQNT.
Basic research in radio astronomy at the Naval Research Laboratory is supported by 6.1 Base funding. The National Radio Astronomy Observatory is a facility of the National Science Foundation operated under cooperative agreement by Associated Universities, Inc. We thank the staff of the GMRT, who have made these observations possible. GMRT is run by the National Centre for Radio Astrophysics of the Tata Institute of Fundamental Research. 
ACF is supported by ERC Advanced Grant 340442



\bibliographystyle{mnras}
\bibliography{biblio_perseus_rev}






\bsp	
\label{lastpage}
\end{document}